\pdfoutput=1
\documentclass[journal]{IEEEtran}
\usepackage[boxed,vlined]{algorithm2e}
\usepackage{graphics}

\SetKw{KwProcedure}{procedure}
\SetKw{KwDownto}{downto}
\SetKw{KwFunction}{function}
\SetKw{KwReturn}{return}

\newtheorem{theorem}{Theorem}[section]
\newtheorem{lemma}{Lemma}[section]

\begin{document}

\title{RBO Protocol: Broadcasting Huge Databases for Tiny Receivers}

\author{Marcin~Kik\thanks{M. Kik is with the 
Institute of Mathematics and Computer Science
of Wroclaw University of Technology, ul. Wybrzeze Wyspianskiego 27
50-370 Wroclaw, Poland
 (e-mail: Marcin.Kik@pwr.wroc.pl).}%
\thanks{This work was supported by MNiSW grant N N206 1842 33.}}

\IEEEcompsoctitleabstractindextext{%
\begin{abstract}
We propose a protocol (called RBO) 
for broadcasting 
long streams of single-packet messages 
over radio channel for tiny,
battery powered,
receivers.
The messages are labeled by the keys from some linearly ordered set.
The sender repeatedly broadcasts a sequence of many (possibly millions) of messages,
while each receiver is interested in reception of a message
with a specified key within this
sequence.
The transmission is arranged so that the receiver can wake up in arbitrary moment and find
 the nearest transmission of its searched message.  
Even if it does not know the position of the message in the sequence,
it needs only
to receive a small number of  (the headers of) 
other messages to locate it properly.
Thus it can save energy by keeping the radio switched off most of the time.
We show that bit-reversal permutation has ``recursive bisection properties''
and, as a consequence,
RBO can be implemented very efficiently 
with only constant number of $\lceil\log_2 n\rceil$-bit variables,
where $n$ is the total number of messages in the sequence. 
The total number of the required receptions is at most $2\lceil\log_2 n\rceil+2$ 
in the model with perfect synchronization.
The basic procedure of RBO 
(computation of the time slot for the next required reception)
requires only $O(\log^3 n)$ bit-wise operations.
We propose implementation mechanisms 
for realistic model (with imperfect synchronization),
for operating systems (such as e.g. TinyOS).
\end{abstract}

\begin{IEEEkeywords}
Radio network, broadcast scheduling, energy efficiency.
\end{IEEEkeywords}}

\maketitle

\IEEEdisplaynotcompsoctitleabstractindextext

\IEEEpeerreviewmaketitle

\section{Introduction}
Recursive Bisection Ordering (RBO) Protocol
is a protocol,
based on a very simple ranking 
algorithm \cite{DBLP:conf/adhoc-now/Kik08},
for a powerful sender and energetically tiny 
receivers.
The sender repeatedly broadcasts a sequence of  
messages.
Each message is labeled by a {\em key}.
The time intervals between subsequent starts of message 
transmissions in the sequence are equal. 
We call them {\em time slots}.  
At arbitrary time moment the user of RBO 
(i.e. some application running on the receiver device)
may ask the RBO module to receive a message with some specified key.
Since then, the task of the RBO module is to receive the nearest transmission 
of the  message labeled with this key and deliver this message to the user.
The simplest strategy would be keeping the radio switched on and listen to all
messages until the searched one is received.
However, radio consumes a lot of energy while it is switched on and the receiver 
device has a limited energy source (i.e. battery).
If the whole sequence contains millions of messages, then
we may need to wait many hours until the searched message is transmitted.
Therefore we need a strategy that minimizes the total radio working time
and does receive the nearest transmission of the searched message. 


Finding broadcast scheduling that
optimizes energy consumption in the battery powered 
receivers becomes one of the main 
problems in diverse modern applications.
An example is a very recent algorithm
of finding optimal scheduling of broadcast 
bursts for mobile TV channels  \cite{5337931}.

Other example are 
wireless networks of battery powered 
sensors.
Nodes of such network consist of possibly 
simple processor, a very limited memory, 
specialized sensing or measurement tools,
and radio receiver and transmitter.
Usually, the task of such network is
reporting the measurements or detected events to the base station.
The radio receiver can be used for forwarding packets from 
the other more
distant sensors,
since  the range of the sensor's transmitter is in many cases
shorter than the distance to the base station (to save the energy).
The other application of the radio receiver 
can be receiving control messages from the base station.
However, keeping the radio receiver switched on all the time
would consume too much energy. 
Techniques for sensor networks 
such as Low Power Listening (LPL, \cite{LPL}),
where the receiver samples for short periods radio channel
and continues listening if it detects any transmission,
while the sender transmits a sequence of few copies of the message
to ensure one successful reception, 
are appropriate for an extensively used channel.
On the other hand, 
RBO is appropriate for a channel with
continuous stream of messages,
where each receiver wants to receive only few of them. 
Also the sleeping intervals for LPL are constant
(and so are the energy savings),
while RBO flexibly adapts the sleeping intervals.
They can be very long for very long sequences of messages.

RBO can be used for transmission of public large databases 
that can be accessed by battery powered devices such as palm-tops.
However, the efficiency and simplicity of its implementation
makes it also useful for very weak devices such as sensors.
For example, it enables sending control commands 
to a great multitude of sensors over a single radio channel.
Each receiver can use RBO to filter its own messages without any prior
knowledge about the transmission schedule.
In such system, we can add/remove receivers without affecting the 
behavior of the other receivers. 
Thus, we have a simple and flexible mechanism for time-division
multiplexing of messages on a single radio channel.
Note that in future we may face the problem of 
broadcasting of a very large amounts of information 
to multitude of energy constrained devices 
scattered in our solar system.
The only transmission medium would be limited number of radio channels.

Another application of the RBO can be centralized channel access
control for upload transmissions 
(e.g. for overcrowded channel): 
The base station broadcasts only the headers,
while the rest of the time slot can be used for
transmission by the (unique) owner of the key from the header.
It can also be considered for broadcasting 
interrogation signals for reporting selective readings 
from sensors or battery powered (gas/water) meter 
devices. 
This could be generalized to the idea of 
distributed algorithms performed by 
sensors (such as e.g. routing towards the base station)
assisted by a powerful base station 
broadcasting
control/synchronization commands organizing the distributed computation.


Transmitting large database for battery powered
receivers has been considered 
by Imielinski, Viswanathan and  Badrinath
in 
\cite{DBLP:conf/sigmod/ImielinskiVB94},
\cite{DBLP:conf/edbt/ImielinskiVB94},
and 
\cite{DBLP:journals/tkde/ImielinskiVB97}.
They proposed several techniques based on hashing and inclusion of
indexing informations in data stream that let the receiver 
energetically efficient searching for data. 
  
Specific variants of the problem and efficiency measures
have also been considered:
Broadcast scheduling minimizing latency in the presence of errors
has been considered in \cite{DBLP:journals/winet/VaidyaH99}.
In \cite{DBLP:journals/isci/LeuH07} data-caching for energy saving
has been proposed.
Energy efficient indexing for 
for several types of data formats
has been proposed in 
\cite{DBLP:journals/tkde/ChungYK10},
\cite{DBLP:journals/isci/ChungL07},
\cite{DBLP:journals/isci/Chung07}.

We believe that, in many applications, 
RBO can be a more implementable and robust solution.
In RBO, each message, consisting of the header and data field,
is of the same type, and occasional losses of messages
do not cause severe consequences.

The RBO protocol is based on a simple ranking algorithm
for single hop radio network proposed in \cite{DBLP:conf/adhoc-now/Kik08}.
The sender sorts the messages by their keys and then permutes them by a special permutation
(called {\em recursive bisection ordering} or {\em $rbo$}).
Such sequence is periodically broadcast.
The receivers' RBO protocol 
keeps an interval $[minR, maxR]$ of possible ranks of the searched key 
in the transmitted sequence.
Initially $[minR, maxR]=[0, n-1]$, where $n$ is the length of the sequence.
RBO tries to receive {\em only} the messages with the keys ranked in $[minR, maxR]$.
Each such message is either the searched one
or it can be used for further updating (shrinking) of the interval.
It has been shown in \cite{DBLP:conf/adhoc-now/Kik08} that 
no more than $4\lg_2 n$
messages are required to locate the rank of the
key in the sequence if the sequence is retransmitted in rounds,
even when the search is started in arbitrary time slot.

In this paper we show that a simple bit-reversal permutation
(famous for its application in FFT \cite{CormenLR89})
has the essential ``recursive bisection'' properties of the 
(recursively defined) $rbo$.
This enables very efficient and simple implementation
of the functions needed by the RBO protocol.
Hence, RBO can be implemented on very weak devices with tiny memory
resources (such as e.g. sensors).

In section~\ref{preliminaries-section}
we show the properties of bit-reversal permutation
that are relevant for our protocol.
We also present the outline of the underlying
algorithm.

In section \ref{reliable-efficiency-section}
we show precise upper bound on
the number of necessary receptions required 
to reach the searched message.
The bound is $2\lceil\lg_2 n\rceil+2$.
Due to the simpler permutation and more detailed proof,
this bound is lower than the one in \cite{DBLP:conf/adhoc-now/Kik08}.
We show an example, when $2\lceil\lg_2 n\rceil-1$ receptions are required.
We also include experimental results of the simulations,
in the case when the communication is unreliable.

In section \ref{nextSlotIn-section} we propose simple and efficient 
 algorithm
for computing the time-slot of the next message that should be received by the receiver.
The algorithm enables 
computations for very long sequences of messages (possibly many millions or more) 
even on very weak processors. 
It requires $O(\log^3 n)$ bit-wise operations and a constant number
of $\lceil\log_2 n\rceil$-bit variables.

In section \ref{implementation-section}
we discuss  
the implementation of the protocol on real devices.
A prototype of the protocol with a simple demonstration application 
has been implemented in Java language
and is available at \cite{RBO-WWW}.
This implementation is designed to be 
easily transformable to TinyOS (\cite{TinyOSProgramming}, \cite{TinyOS-WWW}):
the required modules of TinyOS, hardware components and radio channel
has been modelled by appropriate objects.
RBO protocol offers split-phase interface to the user.
The user issues a command to find a message with given key and,
after some time is signalled the call-back with the results
of the search.
In the meantime RBO switches the radio receiver on and off:
on the one hand -- to save energy, on the other hand -- 
to ensure the reception of all the messages required for the search.
Also the basic protocol functions have been implemented  
with no recursion
and optimized up to the bitwise operations.

\section{Preliminaries and Relevant Properties of Bit-reversal}\label{preliminaries-section}

There is a single {\em broadcaster} and arbitrary number of receivers.
The broadcaster has a set of $n$ {\em messages} to be broadcast labeled by {\em keys}
from some linearly ordered universe. 
The keys do not have to be distinct.
The broadcaster sorts the messages by the values of their keys.
By a {\em rank} we mean a position index of an item in this sorted sequence.
(The positions are numbered from $0$ to $n-1$.)
Then the broadcaster broadcasts in a round-robin fashion the sorted sequence of messages
permuted by a fixed permutation $\pi$, i.e.:
the message with rank $x$ is broadcast in the time slots that are congruent modulo $n$ to $\pi(x)$. 
On the other hand, each receiver can at arbitrary time slot start the Algorithm~\ref{algorithm-receiver} 
described below (technical re-formulation of ranking proposed in \cite{DBLP:conf/adhoc-now/Kik08})
 to receive the message with a specified key.

We assume that the length of the transmitted sequence is $n=2^k$, for
some positive integer $k$.
(If the actual number of messages is not a power of two, then
 we can duplicate some of them to obtain a sequence of length $2^k$.)

For $k\ge 0$ and $x\in\{0,\ldots,2^k-1\}$
we define:
$$
revBits_k(x)= \sum_{i = 0}^{k -1} 2^i \cdot x_{k - 1 - i},
$$
where $x_i = \lfloor x / 2^i\rfloor \bmod 2$.
Note that if 
$(x_{k-1},\ldots,x_0)_2$ is a binary representation of $x$,
then $(x_{0},\ldots,x_{k-1})_2$ is  a binary representation of 
$revBits_k(x)$.
We call  $revBits_k$ a {\em $k$-bit-reversal permutation}.

We argue, that bit-reversal is a good choice, 
for the permutation $\pi$ mentioned
above, for the following reasons:
\begin{itemize}
\item
The low energetic costs of the radio operation of the receiver
(see Section~\ref{reliable-efficiency-section}).
\item 
The simplicity and efficiency of the implementation of the function 
$nextSlotIn$
(see Section~\ref{nextSlotIn-section} and \cite{RBO-WWW}).
\item 
Also the results of simulations (see Figure~\ref{wykres-fig})
show the robustness of the algorithm to 
random loses of messages, e.g. caused by  external interferences. 
\end{itemize}

\begin{figure}
\centering{%
\resizebox{!}{35mm}{
    \includegraphics{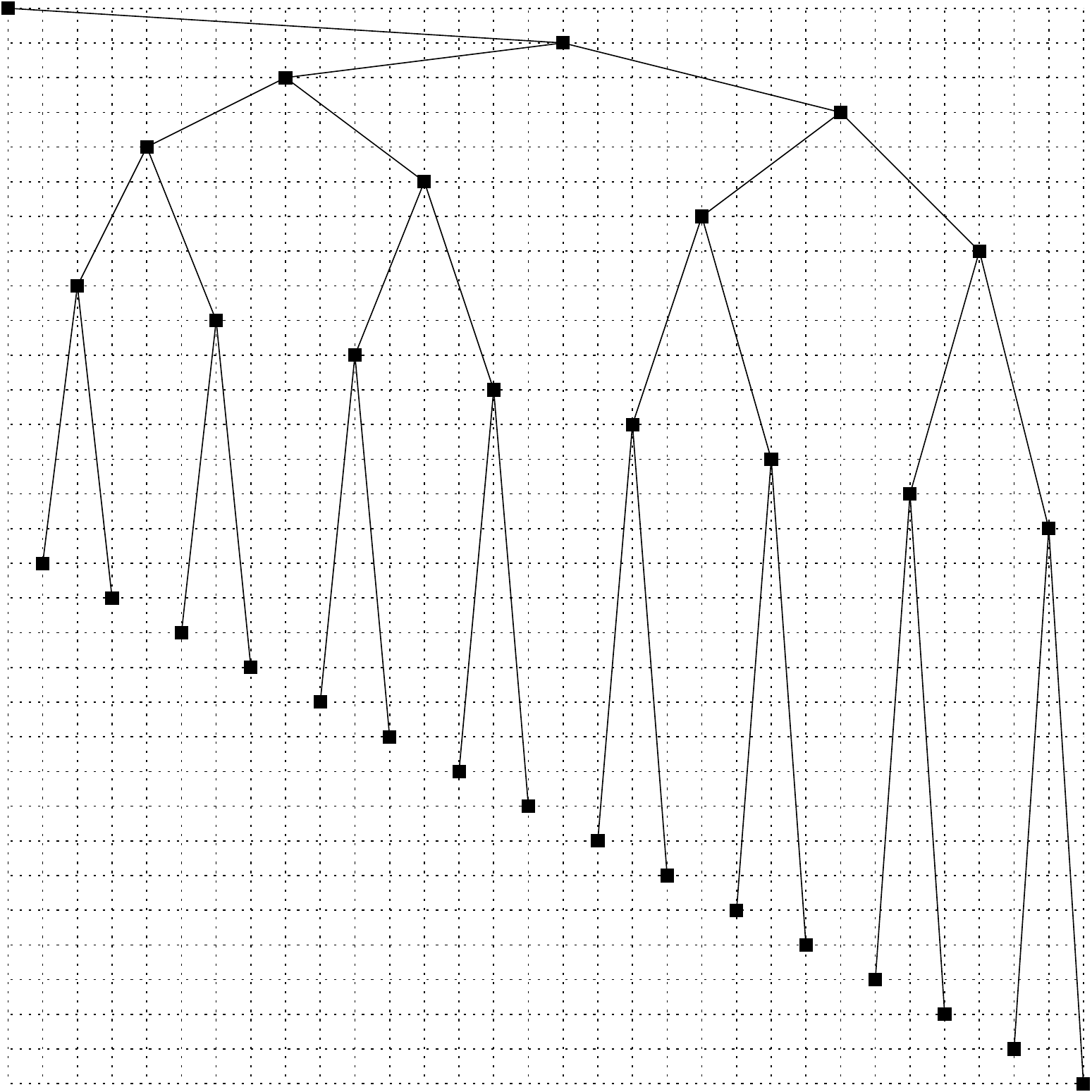}\hspace{2cm}%
    \includegraphics{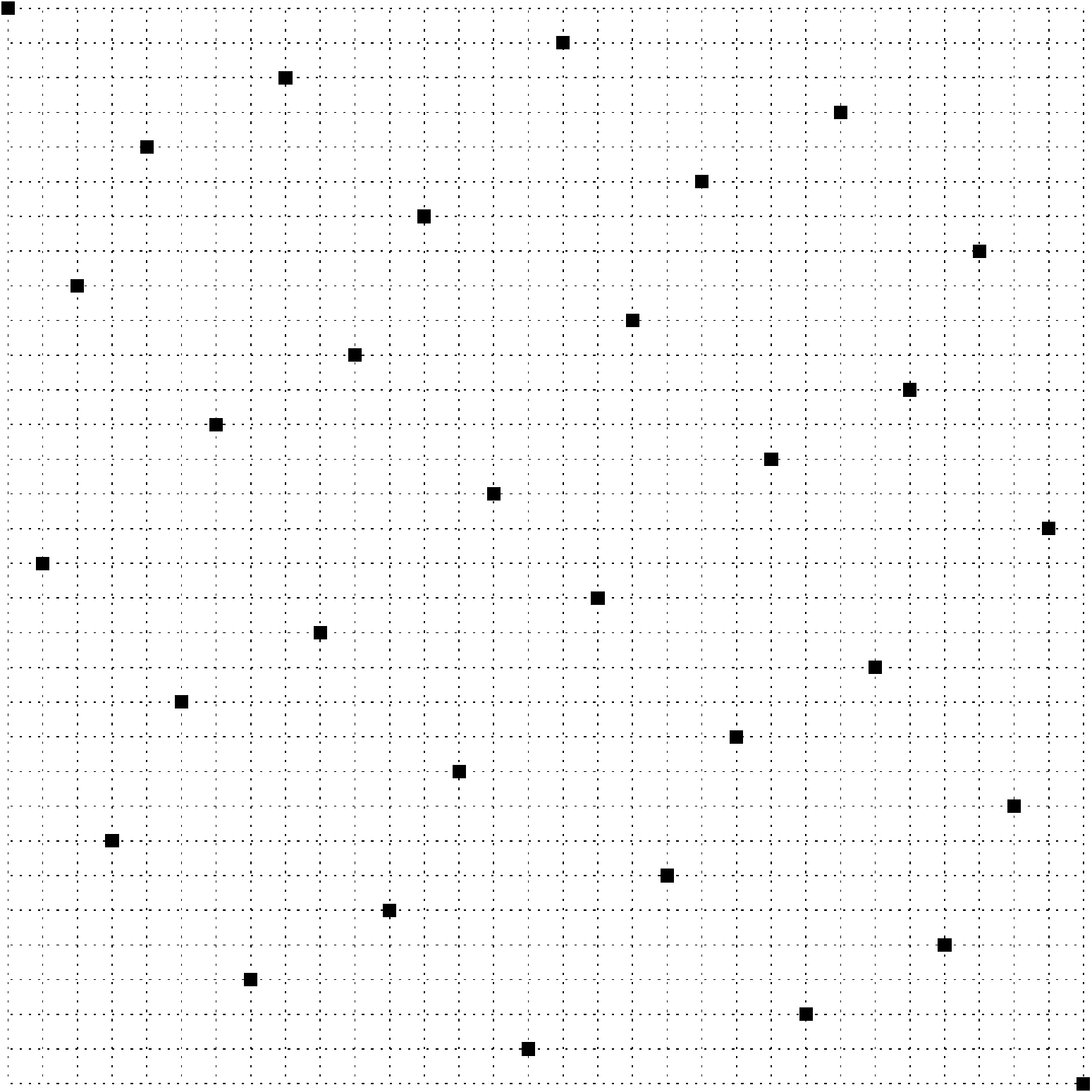}
  }\\
\vspace{5mm}
\resizebox{!}{35mm}{
    \includegraphics{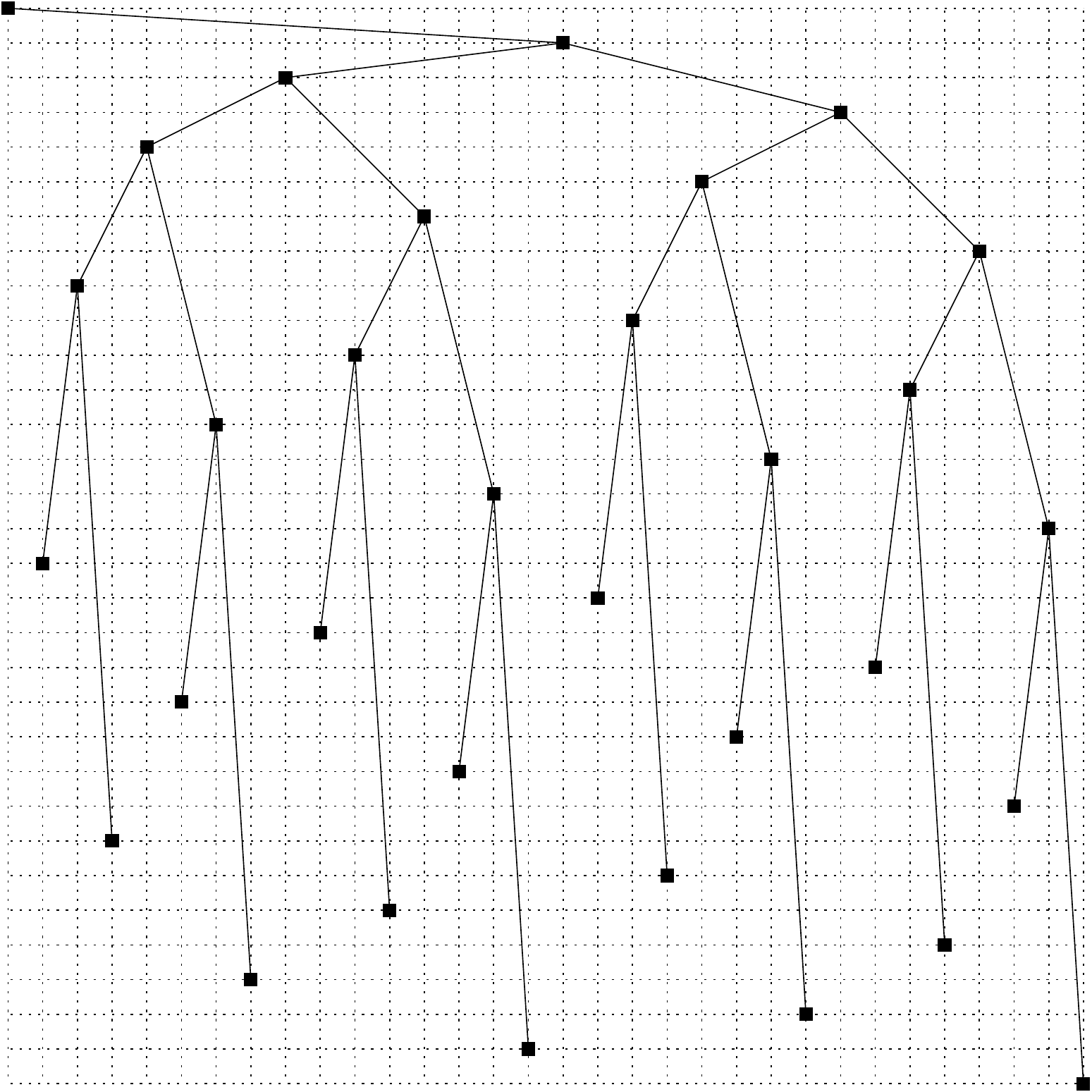}\hspace{2cm}%
    \includegraphics{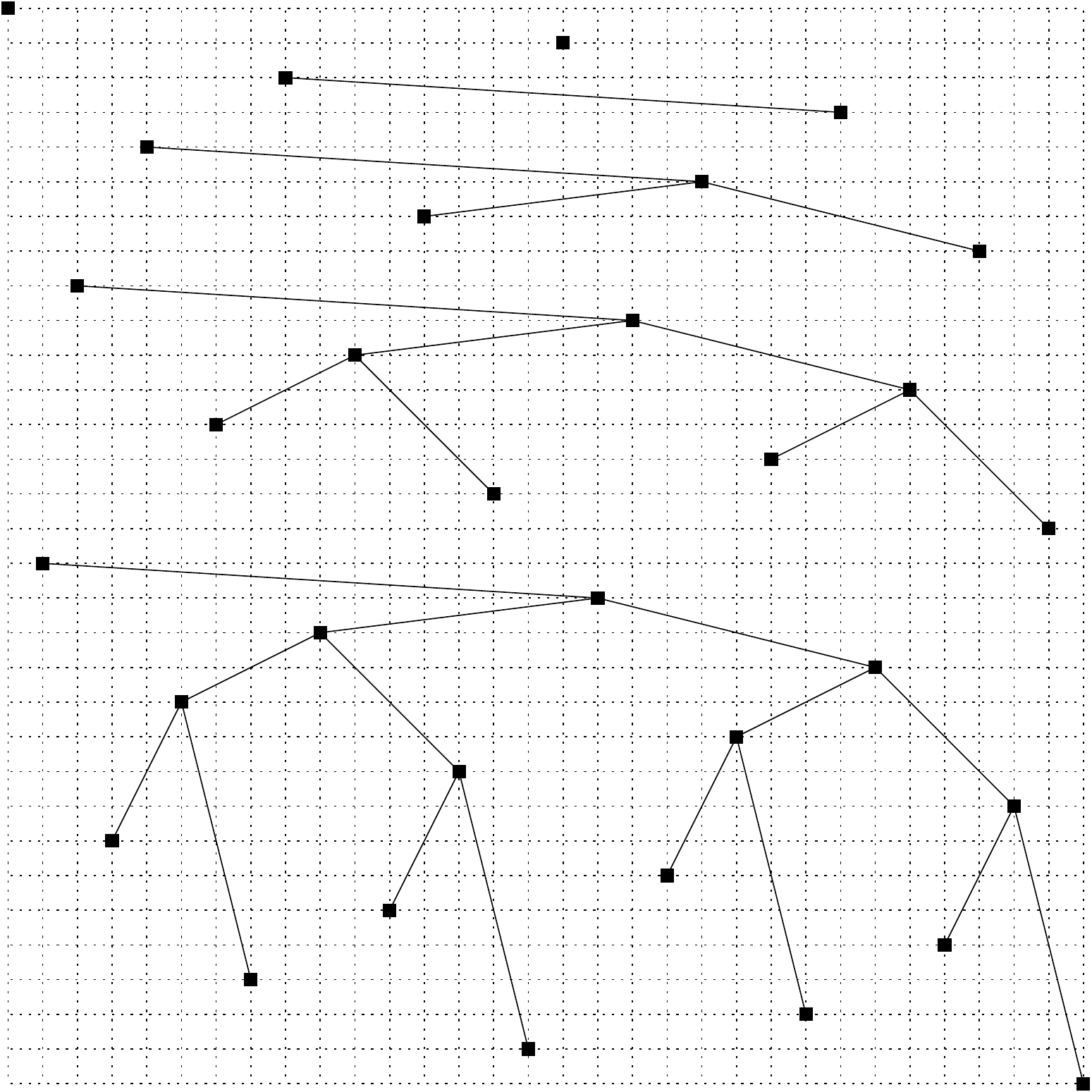}
  }
}
\caption{The graphs of permutations: 
$bs_5$ with embeded tree (upper left),  
$revBits_5$ (upper right),
$revBits_5$ with embeded tree (lower left), 
and with the trees on the first recursion level (lower right).
On the graphs, 
the axis of domain (corresponding to {\em ranks}) is directed rightwards, 
while the axis of the range (corresponding to {\em time}) is directed downwards.%
\label{fig-trees}}
\end{figure}

A natural efficient solution to the problem of finding a key in the sorted
sequence is application of the {\em binary searching}.
We can define an (almost) balanced binary search tree on $2^k$ nodes.
As the first approach we define a permutation $bs_k$ 
(see the upper left graph on Figure~\ref{fig-trees}).
For $k\ge 0$, let $bs_k$ ({\em binary search ordering})
be a permutation of $\{0,\ldots,2^k-1\}$ defined as follows:
\begin{itemize}
\item ${bs}_0 (x)=0$, and,
\item ${bs}_{k+1} (x)  =  (1-(x\bmod 2))\cdot {bs}_{k} (\lfloor x / 2 \rfloor) 
                            + (x\bmod 2) \cdot (2^{k} + \lfloor x / 2 \rfloor)$.
\end{itemize}
The domain of the permutation corresponds to {\em ranks}, 
while its range corresponds to {\em time slots}.
In the definition of $bs_{k+1}$,
for each even rank $x$, only the component:
``$(1-(x\bmod 2))\cdot {bs}_{k} (\lfloor x / 2 \rfloor)$'' can be non-zero,
and, for each odd rank $x$, only the component:
``$(x\bmod 2) \cdot (2^{k} + \lfloor x / 2 \rfloor)$'' can be non-zero.
Thus, all the even ranks, permuted by $bs_{k-1}$ (ignoring the least significant -- parity -- bit), 
are placed before 
the odd ones -- the {\em leaves} of {\em binary search tree}.
Th upper-left graph on Figure~\ref{fig-trees} is the graph of $bs_k$ for $k=5$.
The axis of the range (the vertical axis) is directed {\em downwards}.
The line segments form the  binary search tree.
A node $(x,y)$ on the graph is on the level $\lceil\lg_2(y+1)\rceil$ of the binary search tree. 
If the sender transmits a sorted sequence of length $2^k$ permuted by $bs_k$
and the receiver starts listening in time slot zero, then it needs to 
receive no more than $k$ keys to locate its searched key.
However, if the receiver starts at arbitrary time, then it may be forced to 
receive many messages.
(Consider the case, when the receiver starts in time slot $2^{k-1}$ and the searched key
is greater than all the keys of the sequence.)
In {\em binary search} it is essential, 
that all the nodes from one level precede all the nodes from the next level.
However, the ordering of the nodes within each level may be arbitrary.  
Note that $revBits$ satisfies the following recurrences:
\begin{itemize}
\item   ${revBits}_0(x) =  0$, for $x=0$, and,
\item   ${revBits}_{k+1}(x)  =  {revBits}_{k} (\lfloor x / 2 \rfloor) 
                         + (x\bmod 2) \cdot 2^{k}$, for $0\le x\le 2^{k+1}-1$.
\end{itemize}
In the definition of ${revBits}_{k+1}$,
both the set of odd ranks (mapped to the time slots $\{2^k,\ldots,2^{k+1}-1\}$ -- the leaves) and the set of the even ranks 
(mapped to the time slots $\{0,\ldots,2^k-1\}$ -- the part of the tree above the leaves)
are both permuted recursively by 
${revBits}_{k}$ (ignoring the parity bit) within their ranges of time slots.
Hence, by the recursion, each level $l$ (of size $\lceil 2^{l-1}\rceil$) is 
{\em recursively permuted} by $revBits_{l-1}$.
The nodes within level $l$ of the binary search tree form a binary search tree and
the same holds for the sub-levels of the level.
The binary search tree for $revBits_5$ and the trees for its levels (the first level of recursion) 
are  shown on the graphs on Figure~\ref{fig-trees}.

The {\em binary search tree} of $revBits_k$ has $k+1$ {\em levels}: $0$,$\ldots$,$k$.
By the {\em level} of the {\em time slot} $t$ we mean $\lceil \lg_2 (t+1)\rceil$,
and by the {\em level} of the {\em rank} $x$ we mean $\lceil \lg_2 (revBits_k(x)+1)\rceil$.
For each rank $x$ on level $l$, we have $0\le x<2^k$ and $x=2^{k-l}+i_x\cdot 2^{k-l+1}$, for some integer $i_x$
called {\em coordinate of $x$ within level $l$}.
Note that $i_x=\lfloor x/2^{k-l+1}\rfloor$.

We use notation $(a_1,a_2,\ldots, a_m)$
to denote a sequence of the elements $a_1$,$a_2$, $\dots$, $a_m$.
Thus, $()$ denotes an empty sequence.
For sequences $\alpha_1$ and $\alpha_2$,
let $\alpha_1\cdot\alpha_2$ denote the 
concatenation of $\alpha_1$ and $\alpha_2$.
Let $|\alpha|$ denote the length of the sequence $\alpha$.
For a decreasing sequence $\alpha$ of numbers from $\{0,\ldots,k\}$, 
we define the set $Y^k_\alpha$ as follows:
\begin{enumerate}
\item
  $Y^k_{()}=\{0,1,\ldots, 2^k-1 \}$.

\item for $0\le l\le \lg_2 |Y^k_\alpha|$,
  $Y^k_{\alpha\cdot (l)}=\{ y\,|\, \lceil \lg_2(y-\min Y^k_\alpha+1)\rceil = l \}$.
\end{enumerate}
We use $Y^k_\alpha$ to denote the subsets of time slots.
$Y^k_{()}$ is the set of all the time-slots 
and $Y^k_{\alpha\cdot (l)}$ 
is the set of time slots 
on the $l$th level of the binary search tree $Y^k_\alpha$.
The following properties are simple consequence of the definition:
\begin{lemma}\label{basic-lemma}
\begin{enumerate}
\item
  $|Y^k_{\alpha\cdot(0)}|=1$ and,
  for $0<l\le \lg_2 |Y^k_\alpha| $, $|Y^k_{\alpha\cdot(l)}|=2^{l-1}$.
\item
  $Y^k_\alpha$ is
  a disjoint union of the sets $Y^k_{\alpha\cdot (l)}$,
  where $0\le l\le \lg_2 |Y^k_\alpha|$. 
\item
$y\in Y^k_{\alpha\cdot (l)}$
if and only if
$\min Y^k_{\alpha}+\lfloor 2^{l-1}\rfloor \le y< \min Y^k_{\alpha}+2^l$.
\item\label{basic-minimum}
  $\min Y^k_{(l_0,l_1,\ldots,l_r)}=\sum_{i=0}^r \lfloor 2^{l_i-1}\rfloor$.
\end{enumerate}
\end{lemma}
Let $X^k_\alpha=revBits_k(Y^k_\alpha)$ 
-- the set of the {\em ranks}
of the time slots $Y^k_\alpha$.

Let us define $step^k_\alpha$ as follows:
\begin{itemize}
\item
if $\alpha=()$ then $step^k_\alpha = 1$, else
\item
if $\alpha=\alpha'\cdot(l)$ then $step^k_\alpha = 2^{k-l+1}$.
\end{itemize}
\begin{lemma}\label{step-lemma}
For each 
$X^k_\alpha$,
we have:
\begin{enumerate}
\item\label{round-robin}
$x\in X^k_\alpha$ if and only if $x=(\min X^k_\alpha+i\cdot step_\alpha)\bmod 2^k$,
for some integer $i$.
\item\label{step-lemma-min}
$step_\alpha\ge \min X^k_\alpha+1$.
\item\label{step-lemma-max}
$\max X^k_\alpha+ step_\alpha\ge 2^k$.
\end{enumerate}
\end{lemma}
\begin{IEEEproof}
If $\alpha=()$, then $X^k_\alpha=\{0,1,\ldots, 2^k-1\}$ and the lemma follows.
Otherwise, $\alpha=\alpha'\cdot(l)$, for some $\alpha'$ and $l$.
If $l\in\{0,1\}$, then $|X^k_\alpha|=1$, $step_\alpha\in\{2^{k+1},2^{k}\}$ and 
the lemma follows.
Otherwise,
 $y\in Y^k_\alpha$ if an only if $\min Y^k_{\alpha'}+2^{l-1}\le y<\min Y^k_{\alpha'}+2^l$.
Note that $\alpha$ is decreasing and,
by Lemma~\ref{basic-lemma}(\ref{basic-minimum}), 
$\min Y^k_\alpha=\min Y^k_{\alpha'}+2^{l-1}$ is divisible by $2^{l-1}$.
In other words:
$Y^k_\alpha$ (respectively, in $X^k_\alpha$) 
is the set of all the numbers that
have the ${k-l+1}$
most significant (respectively, least significant) bits
identical to $\min Y^k_{\alpha}$ (respectively, $revBits_k(\min Y^k_{\alpha})$). 
Hence, 
$X^k_\alpha=revBits_k(Y^k_\alpha)=$ 
$\{x\,|\, 0\le x< 2^k\,$
$\wedge\, x\bmod 2^{k-l+1}=revBits_k(\min Y^k_{\alpha})\}$.
Since 
$step_\alpha=2^{k-l+1}$, we have
\begin{itemize}
\item
$step_\alpha>revBits_k(\min Y^k_{\alpha})=\min X^k_\alpha $ and 
\item
$x\in X^k_\alpha$ if and only if $x=(\min X^k_\alpha+i\cdot step_\alpha)\bmod 2^k$,
for some integer $i$.
\end{itemize}
Thus the lemma follows.
\end{IEEEproof}

Notice that, for $l>0$, $x\in X^k_{\alpha\cdot(l)}$ if and only if 
$x-step^k_{\alpha\cdot(l)}/2 \in  \bigcup_{i=0}^{l-1} X^k_{\alpha\cdot(i)}$.
(The ranks from the level $X^k_{\alpha\cdot(l)}$ are {\em equidistantly interleaved} 
with the ranks from all previous levels $\bigcup_{i=0}^{l-1} X^k_{\alpha\cdot(i)}$.)
Thus:
\begin{lemma}\label{step-lemma2}
For $l\ge 0$, 
$x\in\bigcup_{i=0}^{l} X^k_{\alpha\cdot(i)}$ if 
and only if 
$x=(\min \bigcup_{i=0}^{l} X^k_{\alpha\cdot(i)}+ j\cdot 2^{k-l})\bmod 2^k$,
for some integer $j$.
\end{lemma}

For integer $x$ and set of ranks $X$, 
let $\delta(x,X)= \min(\{\infty\}\cup \{d>0\,|\, x+d\in X\})$,  
and, for non-empty $X$,
let
$minStep(X)=\min(\{ \delta(x,X)\,|\, x\in X\})$.
If $X$ is a singleton, then $minStep(X)=\infty$.
From Lemmas~\ref{step-lemma} and \ref{step-lemma2},
we have:
\begin{lemma}\label{minStep-lemma}
\begin{enumerate}
\item\label{minStep-level-lemma}
  $minStep(X_\alpha)\ge step_\alpha$, and
\item\label{minStep-upper-lemma}
  $minStep(\bigcup_{i=0}^{l} X^k_{\alpha\cdot(i)})\ge 2^{k-l}$.
\end{enumerate}
\end{lemma}

We also state the following simple fact:
\begin{lemma}\label{two-hits-lemma}
If $2\cdot minStep(X)> r_2-r_1$, 
then 
$|[r_1, r_2]\cap X|\le 2$.
\end{lemma}

\subsection{Outline of the Protocol}

The most important function used by the RBO protocol is $nextSlotIn_k$ defined,
for $0\le t<2^k$, $0\le r_1\le r_2<2^k$, as follows:
$$
nextSlotIn_k(t, r_1, r_2)=
\left(t+\tau_k(t,r_1,r_2)\right) \bmod 2^k, 
$$
where
$\tau_k(t,r_1,r_2)=\min\{d>0 : r1\le revBits_k( (t+d)\bmod 2^k   ) \le r_2\}$.
(I.e. the number of the next slot after $t$ with rank in $[r_1,r_2]$.)

The sender simply sorts the sequence of messages by the keys and permutes it by the permutation
$revBits_k$. 
Then it repeatedly broadcasts such sequence.
The receiver contains variables 
$minr$ (initiated to $0$), 
$maxr$ (initiated to $n-1$) 
and the searched key $\kappa$.
The underlying algorithm for the receiver is outlined
in Algorithm~\ref{algorithm-receiver}.
\begin{algorithm}
\Repeat{$minr> maxr$}
  {
  receive message $m$;\\
  (* $m$ contains a key $m.\kappa$ and  $m.rank$ -- the rank of $m.\kappa$ *)\\
  \If{$m.\kappa=\kappa$}
    {
     report the found message $m$ and stop;
    }
  \If{$m.\kappa<\kappa \wedge minr\le m.rank$}
    {
     $minr\gets m.rank+1$;
    }
  \If{$m.\kappa>\kappa \wedge maxr\ge m.rank$}
    {
     $maxr\gets m.rank-1$;
    }
    \If{$minr\le maxr$}
       {
         let $t= revBits_k(m.rank)$;
         \\
         sleep (and skip all transmissions) until the time~slot~$nextSlotIn_k(t, minr, maxr)$;
       }
  }
report the absence of $\kappa$;
\caption{Outline of the receiver's algorithm.\label{algorithm-receiver}}
\end{algorithm}
Thus, the interval $[minr, maxr]$ of possible ranks of the searched 
key $\kappa$ shrinks until it becomes empty 
or the searched key is found.
The sleeping periods between subsequent receptions
rapidly increase as the length of the interval decreases.

\section{Bounds on Time and Energy}\label{reliable-efficiency-section}

\begin{theorem}\label{reliable-theorem}
Let $n=2^k$, for some positive integer $k$.
Let $\kappa_0$,$\ldots$,$\kappa_{n-1}$ be a sorted sequence of keys.
Let $\kappa$ be arbitrary searched key , 
let $t_0$ be arbitrary time slot, $0\le t_0<n$, and,
let $minr_0=0$ and $maxr_0=n-1$.
For $i\ge 0$, let $t_{i+1}= nextSlotIn(t_i, minr_i, maxr_i)$, and,
\begin{itemize}
\item
  if $\kappa<\kappa_{revBits(t_{i+1})}$ then $minr_{i+1}=minr_i$ and $maxr_{i+1}=revBits(t_{i+1})-1$,
  else
\item
  if $\kappa>\kappa_{revBits(t_{i+1})}$ then $minr_{i+1}=revBits(t_{i+1})+1$ and $maxr_{i+1}=maxr_i$,
  else 
\item
  $minr_{i+1}=minr_i$ and $maxr_{i+1}=maxr_i$.
\end{itemize}
Let $e=\min\{i>0 \,|\, minr_i\ge maxr_i \vee \kappa_{revBits(t_{i})}=\kappa\}$.
We have:
\begin{enumerate}
\item\label{reliable-energy}
  $e\le 2\lg_2 n+2$, and
\item\label{reliable-time}
  $t_e$ is at most $n$ time slots after $t_0$.
\end{enumerate}
\end{theorem}
\begin{IEEEproof}

Note that $t_1= (t_0+1)\bmod n$,
and $t_1$, $t_2$, $\ldots$, $t_e$ are the reception time slots 
required by the search for $\kappa$ 
started just before $t_1$. 
If $\kappa\in \{\kappa_0,\ldots,\kappa_{n-1}\}$,
then the sequence $(t_1,t_2,$ $\ldots$, $t_{e-1},t_e)$ is
a prefix of the sequence of time slots used for 
searching for some $\kappa'\not\in \{\kappa_0,\ldots,\kappa_{n-1}\}$
with the same rank as $\kappa$.
Therefore we consider only the case: $\kappa\not\in \{\kappa_0,\ldots,\kappa_{n-1}\}$.

Note that $\{ \kappa_{t_1}, \kappa_{(t_1+1)\bmod n},\ldots, \kappa_{(t_1+n-1)\bmod n}\}$
contains all the keys $\kappa_0$,$\ldots$,$\kappa_{n-1}$.
Hence, the bound on time (part~\ref{reliable-time}) is valid.

Now consider the part~\ref{reliable-energy} (the bound on energy).
Let $U$ denote the set of the (used) time slots $\{t_1,t_2,\ldots,t_{e-1},t_e\}$.

Let $T_i$ be the set of all the time slots since $t_1$ until $(t_{i+1}-1)\bmod n$:
$T_0=\emptyset$ and, 
for $1\le i<e$, 
$T_i=\{(t_1+d)\bmod n\;|\; 0\le d< d_i\}$,
where $d_i=\min\{x\ge 0\,|\,t_{i+1}=(t_1+x)\bmod n\}$.
Let $R_i=revBits_k(T_i)$ be the ranks of the time slots $T_i$.
Lemma~\ref{complete-bounds-lemma} follows from the definition of $nextSlotIn_k$ and $minr_i$ and $maxr_i$:

\begin{lemma}\label{complete-bounds-lemma}
The values $minr_i-1$ and $maxr_i+1$ are the most precise bounds on the rank of $\kappa$
from the subset
$R_i\cup\{-1,n\}$:
\begin{enumerate}
\item\label{complete-bounds-lemma-minR}
$minr_i-1=\max \left(\{-1\}\cup \{x\,|\, \kappa_x<\kappa \wedge x\in R_i \}\right)$, and  
\item \label{complete-bounds-lemma-maxR}
$maxr_i+1=\min \left(\{n\}\cup \{x\,|\, \kappa_x>\kappa \wedge x\in R_i \}\right)$, and 
\item\label{complete-bounds-lemma-delta}
(since $\kappa\not\in\{\kappa_0,\ldots,\kappa_n\}$) 
$maxr_i+1=minr_i-1+\delta(minr_i-1,\{n\}\cup R_i)$.
\end{enumerate}
\end{lemma}

Lemma~\ref{precision-lemma} states that each $Y^k_\alpha\subseteq T_i$
imposes bounds on the length of the interval $[minr_i,maxr_i]$.
\begin{lemma}\label{precision-lemma}
$maxr_i+1\le minr_i-1+\min \{step^k_\alpha\,|\, Y^k_\alpha\subseteq T_i\}$.
\end{lemma}
\begin{IEEEproof}
By Lemma~\ref{complete-bounds-lemma}(\ref{complete-bounds-lemma-delta}),
$maxr_i+1=minr_i-1+\delta(minr_i-1, \{n\}\cup R_i)$.
Let $Y^k_\alpha\subseteq T_i$. 
Since $X^k_\alpha\subseteq R_i$,
we have 
$\delta(minr_i-1, \{n\}\cup R_i)
\le 
\delta(minr_i-1, \{n\}\cup X^k_\alpha)
\le
step^k_\alpha$.
The last inequality follows from Lemma~\ref{step-lemma}:
\begin {itemize}
\item 
if $minr_i-1< \min X^k_\alpha$, then, 
by Lemma~\ref{complete-bounds-lemma}(\ref{complete-bounds-lemma-minR}),
$minr_i-1 \ge -1$ and,
by Lemma~\ref{step-lemma}(\ref{step-lemma-min}),
$\min X^k_\alpha\le step^k_\alpha-1$,
\item 
if $minr_i-1\ge \max X^k_\alpha$, then,
by Lemma~\ref{complete-bounds-lemma}(\ref{complete-bounds-lemma-minR}),
$minr_i-1 < n$ and,
by Lemma~\ref{step-lemma}(\ref{step-lemma-max}),
$n-\max X^k_\alpha\le step^k_\alpha$.
\item
otherwise, 
$minr_i-1$ is between two consecutive elements
in $X^k_\alpha$ which are at the distance $step^k_\alpha$,
by Lemma~\ref{step-lemma}(\ref{round-robin}). 
\end{itemize}
\end{IEEEproof}

Let $\beta$ be the shortest sequence, such that $\min Y^k_\beta= t_1$.
If $\beta=()$, then $t_1=0$ and we start binary search from the global root.
(Thus each of $t_1, \ldots, t_{e}$ is on distinct level and, hence, $e\le k+1$.)
Otherwise, 
let $\beta_0=\beta$ and, for $j\ge 0$, let $\beta_{j+1}$ be defined
as follows:
\begin{itemize}
\item 
  if $\beta_j=()$, then $\beta_{j+1}$ is not defined, 
  else
\item 
  if $\beta_j=\alpha\cdot (l',l, l-1,\ldots, l-m)$,
  where $l+1<l'$ and $m\ge 1$, 
  then $\beta_{j+1}=\alpha\cdot(l',l+1)$, 
  else
\item 
  if $\beta_j=(l,l-1,\ldots,l-m)$,
  where $l<k$ and  $m\ge 1$,
  then $\beta_{j+1}=(l+1)$, 
  else
\item 
  if $\beta_j=(k,k-1,\ldots,k-m)$,
  where $m\ge 0$,
  then $\beta_{j+1}=()$, 
  else
\item 
  $\beta_{j+1}=\alpha\cdot (l+1)$, 
  where $\beta_j=\alpha\cdot (l)$.

\end{itemize}
Let $last=\min\{j\,|\, \beta_j=()\}$.

For $0\le j\le last$, let $f_j$ (the {\em foot} of $\beta_j$) be defined as follows:
\begin{itemize}
\item
if $\beta_j=\alpha\cdot(l)$, for some $\alpha$ and $l$, then let $f_j=l$,
else 
\item 
(i.e. when $\beta_j=()$) let $f_j=k+1$.
\end{itemize}
Note that $f_0>0$, since $\min Y_{\alpha\cdot(0)}=\min Y_{\alpha}$.
The following lemma follows directly from the definitions of $\beta_j$ and $last$.
\begin{lemma}\label{foot-increasing-lemma}
\begin{enumerate}
\item 
$f_0>0$, and
\item
for $0\le j<last$, $f_j+1+|\beta_j|-|\beta_{j+1}|= f_{j+1}$, and 
\item 
$f_{last}= k+1$.
\end{enumerate}
\end{lemma}
Notice that $last\le k$, since $f_0> 0$, and $f_j<f_{j+1}$ (since $|\beta_j|\ge |\beta_{j+1}|$). 

The sequence of time slots $(t_1,(t_1+1)\bmod n,$ $\ldots$ $, t_e)$ is a prefix of the sequence
$\sigma_0\cdot\ldots\cdot\sigma_{last}$, 
where $\sigma_i$ is the sorted sequence of time slots from $Y^k_{\beta_i}$.
Moreover $\sigma_0\cdot\ldots\cdot\sigma_{last-1}$ and $\sigma_{last}$ are
increasing sequences of consecutive integers:
\begin{lemma}\label{Y-beta-decomposition}
\begin{enumerate}
\item
  $\min Y^k_{\beta_0}=t_1$, 
  and
\item
  for $0\le j<last-1$,
  $\max Y^k_{\beta_{j}}+1 = \min Y^k_{\beta_{j+1}}$, 
  and
\item
  $\max Y^k_{\beta_{last-1}}=n-1$, 
  and
\item
  for $0\le i\le last$,
  $\emptyset\not=Y^k_{\beta_{j}} = \{t\,|\,\min Y^k_{\beta_{j}}\le t\le \max Y^k_{\beta_{j}}\}$,
  and
\item
  $Y^k_{\beta_{last}}=\{0,1,\ldots,n-1\}$.
\end{enumerate}
\end{lemma}

We will show the bounds on the sizes of the intersections 
$U \cap Y^k_{\beta_j}$.
\begin{lemma}\label{first-level-lemma}
$|U \cap Y^k_{\beta_0}|\le \lg_2 | Y^k_{\beta_0} |+1=\max\{1,f_0\}\le f_0+1$.
\end{lemma}
\begin{IEEEproof}
$t_1$ is the root of the binary search tree $Y^k_{\beta_0}$ and
the number of levels of this tree is $\lg_2 | Y^k_{\beta_0} |+1=\max\{1,f_0\}\le f_0+1$.
\end{IEEEproof}

Consider the case, when $|\beta_{j}|=|\beta_{j+1}|\ge 1$.
\begin{lemma}\label{next-level-lemma}
If $|\beta_{j}|=|\beta_{j+1}|$ 
then 
$|U\cap Y^k_{\beta_{j+1}}|
\le 
2
\le 
f_{j+1}-f_j+1$.
\end{lemma}
 
\begin{IEEEproof}
We have $\beta_j=\alpha\cdot (l)$ and $\beta_{j+1}=\alpha\cdot (l+1)$,
for some $\alpha$ and $l$.
If $l=0$, then  $|Y^k_{\beta_{j+1}}|=1$.
Otherwise, let  $S=U\cap Y^k_{\beta_{j+1}}$ (time slots used in $Y^k_{\beta_{j+1}}$).
If $S=\emptyset$ then $|U\cap Y^k_{\beta_{j+1}}|=0$.
If $S\not=\emptyset$, then let $s=\min\{i\,|\, t_i\in S\}$.
By Lemma~\ref{Y-beta-decomposition}, we have $Y^k_{\beta_j}\subseteq T_{s-1}$.
By Lemma~\ref{precision-lemma}, $maxr_{s-1}-minr_{s-1}< step^k_{\beta_j}=2\cdot step^k_{\beta_{j+1}}$.
In $Y^k_{\beta_{j+1}}$ we use only the time slots with the
ranks in $[minr_{s-1},maxr_{s-1}]$. Hence $|S|\le |[minr_{s-1},maxr_{s-1}]\cap X^k_{\beta_{j+1}}|$.
By Lemma~\ref{minStep-lemma}(\ref{minStep-level-lemma}), $minStep(X^k_{\beta_{j+1}})\ge step^k_{\beta_{j+1}}$
and, by Lemma~\ref{two-hits-lemma},
$|[minr_{s-1},maxr_{s-1}]\cap X^k_{\beta_{j+1}}|\le 2= (l+1)-l+1=f_{j+1}-f_j+1$.
\end{IEEEproof}

Note that if we have ranked $\kappa$ in 
the levels
$Y^k_{\alpha\cdot(0)},\ldots, Y^k_{\alpha\cdot(l)}$, then
we have to check at most one rank on each level $Y^k_{\alpha\cdot(l')}$ with $l'>l$,
since we simply make a continuation of binary search in the binary search tree $Y^k_\alpha$:
\begin{lemma}\label{continuation-lemma}
If $\bigcup_{i=0}^l Y^k_{\alpha\cdot(i)}\subseteq T_{e-1}$, then,
for each $l'$ such that $l<l'\le \lg_2|Y^k_{\alpha}|$,
we have
$|U\cap Y^k_{\alpha\cdot(l')}|\le 1$.
\end{lemma}

Consider the case, when $|\beta_j|>|\beta_{j+1}|\ge 1$.
\begin{lemma}\label{recursion-back-lemma}
If $|\beta_{j}|>|\beta_{j+1}|\ge 1$ 
then 
$|U \cap Y^k_{\beta_{j+1}}|
\le 
2+|\beta_{j}|-|\beta_{j+1}|
\le
f_{j+1}-f_j+1$.
\end{lemma}

\begin{IEEEproof}
Let $m=|\beta_{j}|-|\beta_{j+1}|$.
Let  $S=U\cap Y^k_{\beta_{j+1}}$.
If $S=\emptyset$ then $|U\cap Y^k_{\beta_{j+1}}|=0$.
If $S\not=\emptyset$, then let $s=\min\{i\,|\, t_i\in S\}$.
By definition, there is a sequence $\alpha$ and 
a level number $l$, such that
$\beta_j=\alpha\cdot (l,l-1,\ldots, l-m)$ and
$\beta_{j+1}=\alpha\cdot (l+1)$.
We split the binary search tree $Y^k_{\beta_{j+1}}$
into upper part $Y'$ and lower part $Y''$ as follows:
Let $Y'=\bigcup_{i=0}^{l-m} Y^k_{\alpha\cdot(l+1,i)}$ and
$Y''=\bigcup_{i=l-m+1}^l Y^k_{\alpha\cdot(l+1,i)}$.
Note that $Y^k_{\beta_{j+1}}=Y'\cup Y''$.
By Lemma~\ref{Y-beta-decomposition}, we have $Y^k_{\beta_j}\subseteq T_{s-1}$.
Let $X'=revBits_k(Y')$.
In $Y'$ we use only the time slots from
$[minr_{s-1},maxr_{s-1}]$, thus
$|U\cap Y'|
\le |[minr_{s-1},maxr_{s-1}]\cap X'|$.
By Lemma~\ref{minStep-lemma}(\ref{minStep-upper-lemma}),
$minStep(X')\ge 2^{k-(l-m)}=step^k_{\beta_j}/2$. 
By Lemma~\ref{precision-lemma},
$maxr_{s-1}-minr_{s-1}< step^k_{\beta_j}$.
Hence, by Lemma~\ref{two-hits-lemma}
we have 
$|[minr_{s-1},maxr_{s-1}]\cap X'|
\le 2$.
Finally, note that,
if $U\cap Y''\not=\emptyset$,
then $Y'\subseteq T_{e-1}$ and, 
by Lemma~\ref{continuation-lemma}, 
$|U\cap Y''|\le m$.
And $2+m=(l+1)-(l-m)+1=f_{j+1}-f_j+1$.
\end{IEEEproof}

For $j<last$, let $c_j= | U\cap Y^k_{\beta_j}|$.
From Lemmas~\ref{first-level-lemma}, 
\ref{next-level-lemma}, and \ref{recursion-back-lemma},
we have:
\begin{lemma}\label{cost-foot-lemma}
$c_0\le f_0+1$, and,
for $0< j<last$, $c_j\le f_j-f_{j-1}+1$.
\end{lemma}

We still need a bound on the number of time slots used since the time slot 0.
Let $U'=\{t\in U\,|\,t<t_1\}$ (equal to $U\setminus \bigcup_{j=0}^{last-1} Y^k_{\beta_j}$).
\begin{lemma}\label{last-bound-lemma}
$|U'|\le k-f_{last-1}+2$.
\end{lemma}
\begin{IEEEproof}
If $U'=\emptyset$ then the lemma follows.
Consider the case $U'\not=\emptyset$:
Let $l=f_{last-1}$.
We split the global binary search tree $Y^k_{()}$
into upper part $Y'$ and lower part $Y''$ as follows:
Let $Y'=\bigcup_{j=0}^{l} Y^k_{(j)}$ and $Y''=\bigcup_{j=l+1}^k Y^k_{(j)}$.
Let $i'=\max\{i\,|\, t_i\ge t_1\}$
(i.e. the index of the last used time slot before the time slot 0).
Let $X'=revBits_k(Y')$.
Since the used time slots in $U'$ have ranks in $[minr_{i'},maxr_{i'}]$,
we have $|U'\cap Y'|
\le |[minr_{i'},maxr_{i'}]\cap X'|$.
Since $Y^k_{\beta_{last-1}}\subseteq T_{i'}$, 
we have, by Lemma~\ref{precision-lemma},
$maxr_{i'}-minr_{i'}<step^k_{\beta_{last-1}}$.
Since, by Lemma~\ref{minStep-lemma}(\ref{minStep-upper-lemma}), 
$minStep(X')\ge 2^{k-l}=step^k_{\beta_{last-1}}/2$,  
we have, by Lemma~\ref{two-hits-lemma},
$|[minr_{i'},maxr_{i'}]\cap X'|
\le 2$.
Finally, note that,
if $U'\cap Y''\not=\emptyset$,
then $Y'\subseteq T_{e-1}$ and, 
 by Lemma~\ref{continuation-lemma}, 
$|U'\cap Y''|\le k-l$.
\end{IEEEproof}

Now, we can  bound $|U|$:
\begin{lemma}\label{energy-lemma}
$|U|\le 2\cdot k+2$.
\end{lemma}

\begin{IEEEproof}
We have $|U|= \sum_{j=0}^{last-1} c_j + |U'|$.
By Lemma~\ref{cost-foot-lemma}, we have
$\sum_{j=0}^{last-1} c_j\le$ 
$f_0+1+\sum_{j=1}^{last-1}(f_j-f_{j-1}+1)=$
$last+f_{last-1}$.
By Lemma~\ref{last-bound-lemma}, we have:
$|U'|\le k-f_{last-1}+2$.
Since $last\le k$, we have
$(last+f_{last-1})+(k-f_{last-1}+2)\le 2k+2$.
\end{IEEEproof}
Lemma~\ref{energy-lemma} completes the proof of Theorem~\ref{reliable-theorem}.
\end{IEEEproof} 

{\em Remark.}
Note that the bound is quite precise:
Consider the case when $\kappa_{n/2}<\kappa<\kappa_{n/2+1}$
and $t_1=\min Y^k_{(2)}$.
Then, on each level $Y^k_{(2)}$,$\ldots$,$Y^k_{(k)}$,
we are using two slots and (in the next round) 
we are using one slot in $Y^k_{(1)}$.
Thus the total number of the used slots is $2(k-1)+1=2k-1$. 

\begin{figure}
\centering{%
\resizebox{!}{40mm}{
    \includegraphics{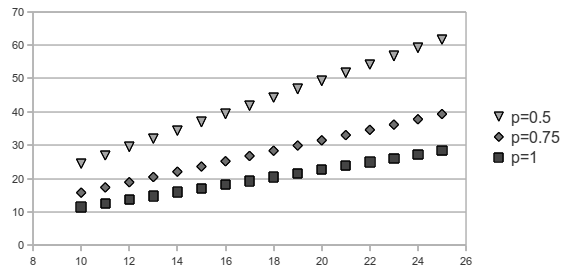}\hspace{2cm}%
  }
}
\caption{Average energy used, for $10\le k\le 25$, for probabilities of successfull
reception $p\in\{0.5, 0.75, 1 \}$.\label{wykres-fig}}
\end{figure}

Theorem~\ref{reliable-theorem} states the bound on energy
under the assumption that every message is received with 
probability $p=1$.
On Figure~\ref{wykres-fig} we present results of 
simulations of the basic algorithm under the assumption
that the probability of successful reception is $p$.
(If the reception is unsuccessful, then a unit of energy is used in
the corresponding time slot, however the interval $[minr, maxr]$
is not updated.)
The horizontal axis is $k$, where $2^k$ is the length of the broadcast sequence,
and the vertical axis is the average energy used by the receiver
in
$100000$
tests. 
In each test,
a random starting time slot $t_0$, $0\le t_0 <2^k$, 
and a key (not present in the broadcast sequence)
with random rank between $0$ and $2^k$ 
have  been uniformly selected.
Since the key is not present in the sequence,
the expected time is bounded
by $(1/p^2-1/2)\cdot 2^k$.

\section{Computation of $nextSlotIn$}\label{nextSlotIn-section}

The function $nextSlotIn_k(t, r_1, r_2)$ is recomputed by RBO 
whenever it has to find the next time slot after $t$, such that
the rank of the key transmitted in this slot is contained
in the interval $[r_1,r_2]$. 
If the rank of the searched key is between $r_1$ and $r_2$,
then RBO can skip all the messages transmitted between time slots $t+1$
and $nextSlotIn_k(t, r_1, r_2)-1$.
Efficient computation of this function reduces the 
time and the energy used by the processor of the receiver device.
If $2^k/(r_2-r_1)$ is not too large (e.g. below one hundred)
then the distance between consecutive elements of $revBits([r_1,r_2])$ 
is not large and we may {\em naively} check sequentially the 
ranks of the time slots $(t+1)\bmod 2^k$, $(t+2)\bmod 2^k$, $\ldots$. 
Otherwise, if $r_2-r_1$ is a small number, then we may apply {\em reverse} searching 
among time slots $revBits(r_1)$, $\ldots$, $revBits(r_2)$, 
for the nearest successor of $t$.
We propose polylogarithmic time computation 
of $nextSlotIn$, 
that should be applied when both
$2^k/(r_2-r_1)$ and $r_2-r_1$ are large.
The implementation of this algorithm in programming language can be found in \cite{RBO-WWW}. 
Here we describe its idea and a more intuitive pseudo-code.
First, let us see how to compute the (globally)  minimal time slot $t$,
such that $revBits_k(t)\in [r_1,r_2]$.
Let $minRevBits_k(r_1,r_2)=\min revBits_k(\{x\, |\, r_1\le x\le r_2\})$.
Note that if $x$ is (the rank of) the node of the binary search tree,
then the left (respectively, right) child of $x$ is $x_L=x-2^{k-l-1}$ 
(respectively, $x_R=x+2^{k-l-1}$), where $l$ is the
level of $x$.
We can compute $minRevBits_k$ 
by following the the path in the binary search tree until
we enter the interval $[r_1,r_2]$ (see Algorithm~\ref{minRevBits}). 
\begin{algorithm}
  \KwFunction $minRevBits_k( r_1, r_2 )$\\
  $x\gets 0$; $s\gets 2^{k-1}$;\\
  \While{$x<r_1$ or $x>r_2$}
      {
        \lIf{$x<r_1$}
            {
              $x\gets x+s$
            }
            \lElse 
                {
                  $x\gets x-s$
                }\\
                $s\gets s/2$;
      }
      \KwReturn $revBits(x)$;
      \caption{Computing $minRevBits$\label{minRevBits}}
\end{algorithm}
By symmetry of $revBits_k$, we have 
that $maxRevBits_k(r_1,r_2)=\max revBits_k(\{x\, |\, r_1\le x\le r_2\})$ is equal to
$2^k-minRevBits_k(2^k-r_2, 2^k-r_1)$.

Here is the outline of our algorithm for computing $nextSlotIn_k(t, r_1,r_2)$:
\begin{enumerate}
\item
If $revBits_k(t)$ is {\em only} one side of the interval $[r_1,r_2]$,
then remove it: 
\begin{itemize}
\item
If $r_1<r_2$ then:
\begin{itemize}
\item
  if $revBits_k(t)=r_1$, then $r_1\gets r_1+1$,
\item
  else if $revBits_k(t)=r_2$, then $r_2\gets r_2-1$.
\end{itemize}
\end{itemize}

\item
If $[r_1,r_2]$ is a singleton then there is no choice:
\begin{itemize}
\item
If $r_1=r_2$ then {\bf return} $revBits_k(r_1)$. 
\end{itemize}

\item  
If $t$ is still before the first slot ranked in $[r_1,r_2]$ in this round,
the return the first slot ranked in $[r_1,r_2]$:
\begin{itemize}
\item
Let $tFirst=minRevBits_k(r_1, r_2)$.
\item
If $t<tFirst$ then {\bf return} $tFirst$.
\end{itemize}

\item 
If $t+1$ is after the last slot ranked in $[r_1,r_2]$,
then 
return the first slot ranked in $[r_1,r_2]$ 
in the next round of broadcasting:
\begin{itemize}
\item
Let $tLast=maxRevBits_k(r_1, r_2)$.
\item
If $tLast\le t$ then {\bf return} $tFirst$. 
\end{itemize}

\item\label{find-level-step}
Here, $tFirst\le t<tLast$.
\begin{itemize}
\item
Find minimal level $l$, such that $l\ge \lceil\lg_2 (t+1)\rceil$ and
$minL=\min \{i\,|\, 2^{k-l}+i\cdot 2^{k-l+1}\ge r_1 \}$
is not greater than
$maxL=\max \{i \,|\,   2^{k-l}+i\cdot 2^{k-l+1}\le r_2 \}$.
\end{itemize}
Such $l$ is the first level (starting from the level of $t$) that
intersects  $[r_1,r_2]$ 
 and
 $\{minL,\ldots,maxL\}$ 
are the 
{\em coordinates  within the level} $l$ of 
this intersection. 
Note that
$minL=\lceil (r_1-2^{k-l})/2^{k-l+1}\rceil = \lfloor (r_2+2^{k-l}-1)/2^{k-l+1}\rfloor$, 
and $maxL= \lfloor (r_2-2^{k-l})/2^{k-l+1}\rfloor$.
The number of nodes above the level $l$
(and also the size of the level $l$) is $2^{l-1}$.
\begin{itemize}
\item
Let $aboveL=2^{l-1}$.

\item 
Let $tFirstL= minRevBits_{l-1}(minL, maxL)$
(the first time slot of the level $l$ ranked {\em within the level} $l$ in $[minL, maxL]$).

\end{itemize}

\item
$aboveL+tFirstL$ is the {\em global} number of the
first time slot of the level $l$
ranked {\em within the level} $l$ in $[minL,maxL]$.
Check whether $t$ is still before this time slot:
\begin{itemize}
\item
  If $t< aboveL+tFirstL$ then {\bf return} $aboveL+tFirstL$.
\end{itemize}

\item
Here $l$ is the level of $t$, since we did not return in previous step.
\begin{itemize}
\item
Let $tLastL= maxRevBits_{l-1}(minL, maxL)$.
\end{itemize}

\item\label{find-level-step1}
If $t\ge aboveL+tlastL$ then 
(we have to find the first slot in $[r_1,r_2]$ below the level $l$): 
  \begin{enumerate}
  \item\label{find-level-step1-substep}
  Find minimal level $l_1>l$, such that 
  $minL_1=\min \{i \,|\, 2^{k-l_1}+i\cdot 2^{k-l_1+1}\ge r_1 \}$
  is not greater than
  $maxL_1=\max \{i \,|\, 2^{k-l_1}+i\cdot 2^{k-l_1+1}\le r_2 \}$.
  ($l_1$ is the next level after $l$ that intersects $[r_1,r_2]$.)
  \item
  Let $aboveL_1=2^{l_1-1}$  (the number of nodes above the level $l_1$).

  \item 
  Let $tFirstL_1= minRevBits_{l_1-1}( minL_1, maxL_1)$.

  \item 
  {\bf Return} $aboveL_1+tFirstL_1$.

  \end{enumerate}

\item 
Here $tFirstL\le t-aboveL <tLastL$ and 
we search within the level $l$ ({\em tail recursion}):
\begin{itemize}
\item
{\bf Return} $aboveL+nextSlotIn_{l-1}(t-aboveL, minL, maxL)$.
\end{itemize}

\end{enumerate}

The depth of the recursion is at most $k$, since each level has no more than 
a half of the nodes of the tree.
Step~\ref{find-level-step1-substep} is 
performed only on the last recursion.
In step~\ref{find-level-step}, $t$ is above level $l$ only on the last recursion.
Thus, the algorithm performs $O(k)$ {\em elementary} operations
such as $revBits$, $minRevBits$, $maxRevBits$ or arithmetic operations. 
Since each such operation  needs $O(k^2)$ bit operations,
the total cost is $O(\log^3 n)$ of bitwise operations.
We replace {\em tail recursion}  by iterative version 
(see the code of \verb|plogNextSlotIn| at \cite{RBO-WWW}). 
Thus RBO uses only constant number of $\lceil\log_2 n\rceil$-bit variables.

\section{Implementation of the Protocol}\label{implementation-section}

We propose an outline of 
practical implementation 
of  RBO for realistic model, 
where the clocks of the sender and of the receiver are
not perfectly synchronized.
We also have to take into account the possible delays
in processing the received messages by the underlying
system protocols.
We have arbitrarily 
selected the set of
available RBO services.
In the case of tiny devices 
such as sensors, 
it is customary that
the code of the
protocol implementation is modified and
tailored to the particular needs 
of the (single) application run on the device.

\subsection{RBO Message Format}
The RBO message consists of  a header and an arbitrary payload.
The header contains the following fields:
\begin{itemize}
    \item 
      $sequenceId$: The identifier of the sequence.
      If sequence of keys changes it should be changed. 
      Zero is reserved for invalid identifier - should not be used.
 
    \item
      $logSequenceLength$: Logarithm to the base of 2 of the sequence length. 
      The length of the sequence is integer power of two.
    \item
      $timeSlotLength$: 
      Time interval between the starts of consecutive message transmissions (e.g. in milliseconds).
    \item
     $key$: The key of the message.

    \item
      $rank$: The rank of the $key$ in the transmitted sequence. 
      Thus the time slot of this message is  $revBits_{logSequenceLength}(rank)$.
\end{itemize}

\subsection{Sender's Part of the RBO}
If the length $n$ of the sequence to be transmitted is not an integer power of two, then
some of the messages should be doubled
 to extend the length to the power of two $n'=2^{\lceil\lg_2 n\rceil}$.
Note that the distance between consecutive occurrences of 
the doubled keys in periodic broadcasting reduces to $2^{\lceil\lg_2 n\rceil-1}$,
while the distance between occurrences of the not doubled keys increases to $2^{\lceil\lg_2 n\rceil}$.
To compensate for this ``injustice'', we can increase the length of the sequence
to even higher power of two by creating  more balanced numbers of copies of the messages.

The sender broadcasts in rounds the sequence of messages sorted by the keys and permuted by the 
$revBits$ permutation.
The messages should have properly filled in header fields.
Whenever the sequence of keys changes, the field $sequenceId$ must be changed unless
 $logSequenceLength$ is changed.

\subsection{Receiver's Part of the RBO}
The RBO module on the receiver's device offers to its user application a {\em split-phase}
interface.
Such interface (see \cite{TinyOSProgramming}) consists of the {\em commands}
to be called by the user and {\em events} to be signalled to the user by the protocol. 
The user (i.e. the running application) issues a command $search(key)$ that initiates
the search and returns immediately.
As soon as the search is finished, the  event call-back $searchDone(message, error)$
is posted to be signaled to the user,
where $message$ is the buffer containing the searched message (if found),
and $error$ is the status of the search result:
\begin{itemize}
\item 
\verb|SUCCESS| (the message has been found),
\item
\verb|KEY_NOT_PRESENT| (the $key$ is not in the sequence),
\item
\verb|TIMEOUT| (no RBO messages has been received for long time),
\item
\verb|BAD_MESSAGE| (an RBO message with $sequenceId=0$ has been received),
\item
\verb|FAILED_RADIO| (problems detected when switching the radio on/off).
\end{itemize}
The user can also pause the current search with the command $stop()$ 
(to be resumed later) 
or abandon it with the command $reset()$ (forgetting all partial results of the search).

On the other hand RBO uses the system modules and interfaces that 
provide the timers ($timeoutTimer$, $sleepingTimer$), 
and the means (e.g. delivered by the TinyOS module \verb|ActiveMessageC|) of packet reception (e.g the interface \verb|Receive|) 
and of switching the radio on and off (e.g. the interface \verb|SplitControl|).  
\begin{figure}
\centering{\resizebox{!}{5mm}{
    \includegraphics{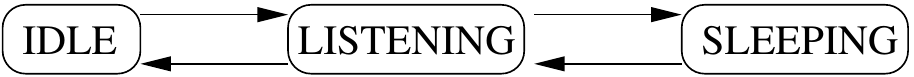}
  }
}
\caption{State diagram of the RBO receiver protocol.\label{RBO-states}}
\end{figure}
RBO can be in one of the three states: 
\begin{itemize}
\item
\verb|IDLE| (when RBO is not used),
\item 
\verb|LISTENING| (when radio is switched on),
\item
\verb|SLEEPING| (when radio is switched off until $sleepingTimer$ fires).
\end{itemize}
The possible state transitions are displayed on Figure~\ref{RBO-states}.
In transition to \verb|LISTENING|,  a $sleepingTimer$ is canceled,  $timeoutTimer$ is set 
and the radio is switched on. 
(Actually, a split-phase process of switching the radio on
is initiated.)
In transition to \verb|SLEEPING|, the $timeoutTimer$ is canceled,
$sleepingTimer$ is set and radio is switched off.
In transition to \verb|IDLE|, the timers are canceled.

RBO has following variables: 
\begin{itemize}
\item
$searchedKey$ -- the recently searched key,
\item
$logSequenceLength$ and $sequenceId$ (initiated to zero) -- recently received in RBO message,
\item
$minRank$ and $maxRank$ -- learned lower and upper bound on the 
rank of $searchedKey$.
\end{itemize}

The user's command $search(key)$ compares $key$ to $searchedKey$
and initiates searching:
\begin{itemize}
\item
If $key<searchKey$, then set $minRank$ to zero.
\item
If $key>searchKey$, then set $maxRank$ to $2^{logSequenceLength}-1$.
\item
Set $searchedKey$ to $key$ and switches RBO to \verb|LISTENING| state.
(Thus we may take advantage from the most recent search.)
\end{itemize}
 
The $stop$ and $reset$ commands switch RBO to \verb|IDLE|.
(Moreover, $reset$ sets $sequenceId$ to zero.)

RBO implements callbacks of the events signalled by the timers and the
interfaces \verb|Receive| and \verb|SplitControl|.
The $timeouTimer$ event $fired()$ in the state \verb|LISTENING|
causes RBO transition to the state \verb|IDLE| and signalling 
$searchDone(\ldots,$ \verb|TIMEOUT|$)$ to the user.
The $sleepingTimer$ event $fired$ in the state \verb|SLEEPING|
causes RBO transition to the state \verb|LISTENING|
and switching the radio on.
The (most essential) event $received( message )$ (reception of the $message$) 
signalled by the radio \verb|Receive|
interface to RBO in state \verb|LISTENING| is served by RBO as follows
(we use notation $message.name$ to denote the field in the $message$ header and $name$ to denote variable of RBO): 
\begin{enumerate}
\item $timeoutTimer$ is canceled.

\item
  If $message.sequenceId=0$,
  then RBO switches to \verb|IDLE| and signals $searchDone($ $message,$ \verb|BAD_MESSAGE|$)$,
  and returns.

\item
  If $message.sequenceId\not=sequenceId$ or $message.logSequenceLength$ $\not=$ $logSequenceLength$, 
  then
  \begin{itemize}
  \item
    set $sequenceId$ to $message.sequenceId$,
  \item
    set $logSequenceLength$ to $message.logSequenceLength$ and
  \item
    (forget old bounds) set $minRank$ to $0$ and $maxRank$ to $2^{k}-1$, where $k=logSequenceLength$.
  \end{itemize}

\item
  If $message.key=searchedKey$ then RBO switches to \verb|IDLE| and signals $searchDone(message,$ \verb|SUCCESS|$)$,
  and returns.

\item Try to update the bounds on the rank:
  \begin{itemize}
  \item
    If $message.key>searchedKey$ and $message.rank\le maxRank$ then set $maxRank$ to $message.rank-1$,
    else
  \item
    if $message.key<searchedKey$ and $message.rank\ge minRank$ then set $minRank$ to $message.rank+1$.
  \end{itemize}

\item
  Test for absence of the $serchedKey$:
  \begin{itemize}
  \item
    If $minRank>maxRank$, then  RBO switches to \verb|IDLE| and signals $searchDone(message,$ \verb|KEY_NOT_PRESENT|$)$,
    and returns.
  \end{itemize}

\item
  Compute the time remaining to the next useful message:
  \begin{itemize}
    \item
      Let $k=logSequenceLength$ and $now=revBits_{k}(message.rank)$ 
      and $next= nextSlotIn_{k}$ $( now, minRank, maxRank)$.
    \item
      If $now<next$ then let $slotsToNext=next-now$, else let $slotsToNext=2^{k}-now+next$.
    \item
      Let $remaingTime=slotsToNext\cdot message.timeSlotLength$.
  \end{itemize}
\item
  If $remainingTime$ is greater then a threshold (i.e. $minSleepingTime$),
  then RBO sets $sleepingTimer$ to $remainingTime-relativeMargin-timeMargin$,
  switches the radio off and transits to state \verb|SLEEPING|,
  where $timeMargin$ is some constant margin (e.g. few milliseconds) 
  that should compensate for radio switching on and off delays and the delay in
  message processing, and $relativeTimeMargin=remainingTime/d$
  should compensate for not ideal synchronization of the sender's and receiver's
  clocks. 
  (If $d$ is a power of two, then the division may be replaced by a binary shift.)
\item
Otherwise  (i.e. when $remainingTime < minSleepingTime$), 
only the $timeoutTimer$ is restarted.

\item
RBO returns.

\end{enumerate}
We skip the descriptions of the implementations of the callbacks $startDone$
and $stopDone$ of the interface \verb|SplitControl| used for switching the radio on and off.
In practice RBO may receive some overhead messages
due to the hardware delays and to keep synchronization with the sender.
The proper balancing of the parameters 
(such as $minSleepingTime$, and the absolute and the relative time margins)
that control the tradeoff between
the energy savings and the reliability can be subject of real life experiments.


\section{Conclusion}

This paper proposes an efficient solution to
the problem of transmitting very long streams 
of uniform messages for 
selective reception by battery powered receivers.

We proposed an implementation of the protocol based 
on a very simple basic algorithm (Algorithm~\ref{algorithm-receiver})
and an efficient algorithm for 
computation of its essential function {\em nextSlotIn}.
Thus, the protocol can be implemented on devices with
very weak processors and with very limited memory.

Note, that we can ``plug-in'' arbitrary permutation
instead of bit-reversal in the basic algorithm.
We have shown 
that,
 for the  bit-reversal permutation,
the number of necessary receptions is bounded by $2\lceil\log_2 n\rceil+2$.
On the other hand we have shown an example,
where $2\lceil\log_2 n\rceil-1$ receptions are necessary.
It is interesting question, whether there exist any
permutation, for which the respective bounds are lower
than for bit-reversal.
However,
$\log_2 n$ is an obvious lower bound and
 the simplicity of bit-reversal
is a great advantage in possible implementations.
%
The tests 
for unreliable 
transmissions (Figure~\ref{wykres-fig})
show that the expected energetic costs are
very low even if the probability of successful reception
is much lower than one.

 \ifCLASSOPTIONcompsoc
   \section*{Acknowledgments}
 \else
  \section*{Acknowledgment}
 \fi

 The author would like to thank Jacek Cicho\'n for helpful comments.

\ifCLASSOPTIONcaptionsoff
  \newpage
\fi

\bibliographystyle{IEEEtran}
\bibliography{kikbib}

\end{document}